\documentstyle[12pt]{article}
\title{
Topological Solitons in Discrete Space-Time \\ as the Model of Fermions}
\author{\large Andrey I. Musienko \\[3mm]
\em N.N. Semenov Institute for Chemical Physics, Russian Academy \\
\em of Sciences, ul. Kosygina 4, Moscow  117977 Russia, \\
\em e-mail: andrew@polymer.chph.ras.ru 
}
\date{}
\begin{document}
\large
\maketitle
\begin{abstract}
In the present paper we discuss arguments, favouring the view that massive
 fermions 
represent dislocations (i.e. topological solitons) in discrete space-time 
with Burgers vectors, parallel to an axis of time.
 If to put symmetrical parts of tensors of distortions ( i.e. derivatives
 of atomic displacements on coordinates) and mechanical stresses equal zero,
 the equations of the field theory of dislocations get the form of the Maxwell 
equations.
 If to consider these tensors as symmetrical, we shall receive the equations of
 the theory of gravitation, and it turns out that the sum of tensor of 
distortions and pseudo-Euclidean metrical tensor  is 
the analogue  of metrical tensor. It is shown that we can also get Dirac
equation with four-fermion interaction in the framework of the field theory of 
dislocations. This model explains quantization of electrical charge: it is 
proportional to the
 topological charge of dislocation, and this charge accepts quantized values 
because of discrete structure of the 4-dimensional lattice.

\end{abstract}

   The concept of discrete (in other words  quantized) space-time 
has been discussed in physics of elementary particles
already for a long time.
 One of variants of this approach is the lattice quantum field 
theory, within the framework of which significant successes were
already achieved.
 It is based on approximation of space-time by a discrete lattice, which is
 actually 4-dimensional analogue of crystal lattices. 
But in the theory of solid state it is well known, that the ideal crystal 
lattices do not exist in a nature.
The real lattices always contain defects, in particular, dislocations.
From the geometrical point of view the dislocations are the one-dimensional 
topological solitons.
The first homotopy group which classifies them is the group of translations.
Therefore it is quite natural to assume, that the similar defects should exist
 in 4-dimensional space-time lattice. Dislocation in (2+1)-dimensional discrete
 space-time with a Burgers vector, parallel to an axis of time, is shown in 
Figure 1. The dislocation line in the figure is parallel to a Burgers vector.

As is known, solitons are particle-like objects; various soliton models of 
elementary particles were offered by many authors. Ross [1] has assumed, that 
fermions represent topological defects in space - time, namely, dislocations 
with Burgers vectors, parallel to an axis of time and taking only quantized 
values. Unzicker [2] offered other topological model of electron: a 
disclination in space-time. The author of paper [2] also used analogy between 
elementary particles and dislocations in crystals. Our approach is closer to 
the hypothesis by Ross [1]. But the consideration by Ross was based solely on 
relation between fermionic spin density  and space-time torsion. In the present
paper we are bringing to your notice a number of other arguments, favouring the
view that fermions represent dislocations in discrete space-time with Burgers 
vectors, parallel to an axis of time. We assume that the discrete space-time 
represents a 4-dimensional lattice. There are some objects ( we shall call 
them "primary atoms" ), capable to be displaced from equilibrium positions, 
in lattice sites. We suppose that the movement of these "primary atoms" obeys 
only the laws of classical Newton mechanics. Certainly, this naive assumption 
is not absolutely justified. It is conceivable that the movement of these 
"primary atoms" can be described by the much more complex laws. But it turns 
out that this "naive" assumption allows by a rather simple way to receive a 
number of the important formulae of electrodynamics, gravitation theory and 
quantum field theory. Probably, account of  deviations of the laws of "primary 
atoms"  behaviour  from the laws of the classical mechanics will allow to 
describe other phenomena (for example, strong or weak interactions).

We shall remind, that torsion is, by definition, an antisymmetric part of 
connection multiplied by two:
$$
T^ {\; \lambda} {} _ {\; \mu \nu} = \Gamma ^ {\; \lambda} {} _ {\; \mu \nu} - 
\Gamma ^ {\; \lambda} {} _ {\; \nu \mu}.
$$
The integral of this function over any 2-dimensional surface, which intersects 
a dislocation, is the value referred to as Burgers vector in solid state 
physics:
$$
b^ {\; \lambda} = \int T^ {\; \lambda} {} _ {\; \mu \nu} \; df^ {\; \mu \nu},
$$
where $ df^ {\; \mu \nu} \equiv dx^ {\; \mu} dx^ {\; \nu} - dx^ {\; \nu} dx^ 
{\; \mu} $ is the element of area.
From the geometrical point of view the space-time, containing dislocations with
 Burgers vectors $b^ {\; i} $ is the manifold with torsion
$T^{\; i}{}_{ah}(x_{\; \zeta })=\frac12 e_{\; achd}\tau ^{\; c} 
b^{\; i}V^{\; d}\delta (x_{\; \zeta }-x_{\; \zeta }^
{\; 0})\;$, where $e_ {\; cahd}
 $ is the 4-dimensional Levi-Civita symbol, $\tau ^ {\; c} $ is the
 4-dimensional unit vector tangential to the dislocation line, $V^ {\; d} $ 
is the 4-dimensional velocity of dislocations, $\delta (x)$ is the Dirac
delta function,
$x_ {\; \zeta} ^ {\; 0} $ are the coordinates  of the dislocation line.
 On the other hand in the theory of gravitation it was shown [3], that the 
fermions
are also sources of torsion.  The relationship between
   a spin density tensor $S^ {\; \lambda} {} _ {\; \mu \nu} $ and a torsion is
 described by the formula [3] 
$$
T^{\; \lambda }{}_{\; \mu \nu }=\frac{8\pi G}{c^3}(S^{\; \lambda }{}_{\; \mu 
\nu }+\frac12 \delta ^{\; \lambda }_{\; \mu }S_{\; \nu }-\frac12 \delta ^{\; 
\lambda }_{\; \nu }S_{\; \mu }).
\eqno (1)
$$
Here 
$G$ is Newton gravitational constant, $c$ is velocity of light,
$S^{\; \lambda }{}_{\; \mu \nu }=v^{\; \lambda }S_{\; \mu \nu }\;$,
$S_{\; \mu }=S^{\; \lambda }{}_{\; \mu \lambda }\;$,
$v^ {\; \lambda} $ is the 4-vector of
    the particle velocity, $S_ {\; \mu \nu} \;$ is the antisymmetric tensor. 
Its spatial components  form a 3-vector $s_ {\; k} = (S^ {\; 23}, S^ {\; 31},
 S^ {\; 12}) $ which is equal to 3-dimensional density of the particle spin   
in reference frame of rest of this particle. In equations (1) and hereinafter 
all the indices, unless otherwise specified, assume the values from 0 to 3.
Therefore, non-moving particle having spin $s_ {\; k} \;$, 
$k=1,2,3$ create the same torsion, as non-moving dislocations with tangential
 vector $\tau _ {\; c} = (\tau _ {\; 0}, \tau _ {\; k}) $ ( vectors 
$s_{\; k}$ and $\tau _{\; k}$ are parallel) and Burgers
 vector, parallel to an axis of time. This dislocation represents a line in 
the 4-dimensional space-time which is coincident with a world line of fermion.
 At any moment of time this line intersects 3-dimensional physical space only 
in one point, therefore fermions are observed as particle objects. Continuity 
of a dislocation line is provided by the topological laws [4]. This fact 
 also was proved with use of the Noether theorem [5]. In continuous space-time
 torsion, created by fixed particles, can accept continuous set of values and 
in quantized space-time it accepts discrete one.
It is well known that spins of elementary particles are quantized. This fact 
results, in accordance with the formula (1), in discreteness of torsion, 
created by fixed particles. 
It confirms hypothesis by Ross [1] of the discrete structure of space-time and 
the dislocation nature of fermions.

To the present time the gauge theory  of dislocations [4-6] is developed very 
well. 
Many authors paid attention on analogy between this theory and both 
electrodynamics [4] and theory of gravitation [7, 8]. 
 There are the equations of incompatibility  in continuous  theory of 
dislocations
$$
\partial _{\; \gamma }\partial _{\; \varepsilon } u_{\; n}(x_{\; \zeta })-
\partial _{\; \varepsilon }\partial
_{\; \gamma }u_{\; n}(x_{\; \zeta })=\frac2c \; T_{\; \gamma n\varepsilon 
}(x_{\; \zeta })
\; ,\eqno  (2)
$$
where $u_ {\; n} $ is the vector of displacement of the 4-dimensional continuum
 particles (in continuous theory the 4-dimensional  lattice is approximated by 
continuous medium ), $\partial
 _{\; \gamma}=\frac{\textstyle \partial }{\textstyle \partial x^{\; \gamma 
}}\; $. Equations (2) are a definition of dislocations and a statement that 
there are no disclinations in the given lattice (since the term descriptive of
the contribution from disclinations is absent in the right-hand side of these 
equations). In other words,  (2) are purely geometrical equations.

 Tensor $\beta _ {\; \varepsilon n} =\partial _ {\; \varepsilon } u_ {\; n} $
 refer to as tensor
 of distortions. Let us introduce tensor $F_ {\; \varepsilon n} $ which equals
 an antisymmetric part of a tensor of distortions multiplied by two:
$$
F_{\; \varepsilon n}=\beta _{\; \varepsilon n}-\beta _{\; n\varepsilon }.
\eqno (3)
$$
By summing each three equations from equations (2), we obtain the following 
system of the equations
$$
\partial _{\; \mu }F_{\; \alpha \beta }+\partial _{\; \alpha }F_{\; \beta \mu
 }+\partial _{\; \beta }F_{\; \mu \alpha }=\frac2c(T_{\; \mu \beta \alpha 
}+T_{\; \beta \alpha \mu }+T_{\; \alpha \mu \beta }).
\eqno (4)
$$
Let the vector tangential to a dislocation line is equal to $\tau _ {\; 
\beta } = 
(\tau _ {\; 0},\tau _ {\; 1}, 0, 0).$ We suppose that the quantities $b_{\; 
0}$ and $\tau _{\; 1}$ are very small. So modern experimental techniques do 
not allow nonzero members to be detected in the right side of (4). Then the 
equations (4) get the form of the first pair of the Maxwell equations
$$
e^{\; \alpha \beta \gamma \delta }\partial _{\; \beta }F_{\; \gamma \delta }=0.
\eqno (5)
$$

Newton second law  for particles of the continuum containing dislocations has 
the form [5]
$$
\partial ^{\; \gamma }\sigma _{\; g\gamma }=\frac2c C_{\; g\varepsilon
 i\lambda }T^{\; \varepsilon i\lambda}
\; .\eqno (6)
$$
The 4-dimensional tensor of mechanical stresses $\sigma_{\; g\varepsilon } $ 
 by definition is equal to 
$$
\sigma_{\; g\varepsilon }=C_{\; g\varepsilon i\lambda }\partial ^{\; \lambda 
}u^{\; i}\; .
\eqno (7)
$$
 Here $C_ {g\varepsilon i\lambda} $ is the 4-tensor of elastic modules of 
4-dimensional continuum. In other words, Lagrangian  of "elastic" waves in 
4-dimensional medium under consideration has the form
$$
{\cal L}_{\; 0}=-\frac12 C^{\; i\gamma j\varepsilon }\beta _{\;
 \gamma i}\beta _{\; \varepsilon j}.
\eqno (8)
$$
By virtue of the fact that the tensor $C_ {g\varepsilon i\lambda} $ is 
sufficiently 
large the distinction from zero of the right side of equations (6) can be
detected in experiments.
 If it is possible to neglect a symmetric part of the tensor  of mechanical 
stresses, the equations (6) get the form of the second pair of Maxwell 
equations. If to consider tensors of distortions and mechanical stresses as 
symmetric, we shall receive the equations  of the gauge translation theory of 
gravitation  (that is theory of gravitation in space-time with zero curvature
and nonvanishing torsion~) [7, 8].
It turns out that the sum of tensor of distortions and the Minkowskian 
(pseudo-Euclidean) metrical tensor is 
the analogue  of metrical tensor in gravitation theory.
 In a general case of dislocations in the 4-dimensional  lattice these tensors 
are neither symmetrical nor antisymmetrical and then we receive as a 
consequence of our model variant of the uniform theory of electrodynamics and
 gravitation, offered by Einstein [9]: tensor of  electromagnetic field is  an 
antisymmetric part of metrical tensor.
 It is simultaneously possible to solve a problem, which  considered by
 Einstein as the main lack of the theory with the asymmetrical metric: to give 
geometrical definition of particles. 

In the theory of dislocations the Frenkel-Kontorova model of a dislocation,
 based on the account
 of only one nonlinear member in Lagrangian of an elastic field, is 
well-accepted [10].
 In the one-dimensional  case it results in the dislocation equation of  motion
 having the form of sine-Gordon equation. The Lagrangian of an 
elastic field in this case has the form
$$
{ \cal L} _ {\; SG} =\frac 12 \partial _ {\; \mu} \phi \; \partial ^ {\; \mu} 
\phi + \frac {m^ {\; 2}} {\beta ^ {\; 2}} [\cos
( \beta \phi) - 1].
\eqno (9)
$$
Coleman [11] has rigorously proved equivalence of the sine-Gordon soliton and 
the fundamental fermion of the massive Thirring model in (1+1) dimensions.
The Lagrangian of the massive Thirring model has the form 
$$
{ \cal L} _ {\; T} = i\bar \psi \gamma _ {\; \mu} \partial ^ {\; \mu} \psi - 
m_ {\; f} \bar \psi \psi -
\frac 12 g (\bar \psi \gamma ^ {\; \mu} \psi)
( \bar \psi \gamma _ {\; \mu} \psi),
\eqno (10)
$$
where $\psi$ is Fermi field, $\gamma ^{\; \mu} $ are Dirac matrices in (1 + 1) 
dimensions. Lagrangians (9) and (10) are equivalent on condition that
$$
\frac {4\pi} {\beta ^ {\; 2}} - 1 = \frac g\pi.
\eqno (11)
$$
 The correspondence between these two models is established by bosonization
 relations:
$$
\frac {m^ {\; 2}} {\beta ^ {\; 2}} \cos (\beta \phi) = - m_ {\; f} \bar \psi 
\psi,
$$
$$
- \frac {\beta} {2\pi} e^ {\; \mu \nu} \partial _ {\; \nu} \phi = \bar \psi 
\gamma ^ {\; \mu} 
\psi \equiv j^ {\; \mu}.
\eqno (12)
$$
The issues of bosonization are considered in more detail in the book by 
Rajaraman [12]. In particular, there was shown that the topological charge 
of the sine-Gordon soliton is equivalent to the fermionic charge of the
 particle 
of the massive Thirring model. Thus, the discreteness of the fermionic charge 
in our approach is a consequence of discreteness of the topological charge of 
solitons (i. e. dislocations), which in turn directly follows from discrete 
structure of space-time.

Many authors consider incompatibility of a space-time lattice with a condition
 of Lorentz invariance
as traditional lack of models of discrete space-time.
 In offered model this contradiction is eliminated.
The Lorentz transformations  arise in this model by a natural way as a 
consequence of finiteness of velocity of light.
 But only values, relating to properties of particles: fields, forces of
 interaction, and so on depend on velocity  by Lorentz law. 
This dependence is a consequence of occurrence of Lorentz roots in expression 
for classical Green function of the equations (6). Therefore quantities, 
expressing through Green function: fields, created by particles, force of 
interaction between particles, and so on only depend on velocity by Lorentz 
law. All other quantities, including 
the parameters of the lattice, are not exposed to any transformations.

In this connection we shall notice that relativistic  expressions always occur
in soliton  theories, in particular, in the theory of dislocations in crystals.
 In this theory instead of velocity of light velocities of sound appear in the 
formulae.
 As in solids even in isotropic case not only transversal, but also
 longitudinal sonic waves can exist, in the theory of dislocations in certain
 cases expressions, containing Lorentz roots of  different kinds: $\sqrt {1-v^ 
{\; 2} /c^ {\; 2} _ {\; \lambda}} \;$, $\lambda=1,2$ can occur. 
For example, in case of straight dislocation  in isotropic medium parallel 
to axis z with Burgers vector $b_ {\; i} = (b, 0,0) $, moving at the velocity
v parallel to axis x, the displacements of particles of continuum  are
 described by the following formulas [13]:
$$
\begin{array}{l}
u_{\; 1}(x,y,t)=bc_{\; 1}^{\; 2}/(\pi v^{\; 2})[
{\rm arctg} (y(1-v^{\; 2}/c_{\; 2}^{\; 2})^{\; 1/2}/(x-vt))+\\ \\
(v^{\; 2}/(2c_{\; 1}^{\; 2})-1)
{\rm arctg} (y(1-v^{\; 2}/c_{\; 1}^{\; 2})^{\; 1/2}/(x-vt))], \\ \\
u_{\; 2}(x,y,t)=bc_{\; 1}^{\; 2}/(2\pi v^{\; 2})[(v^{\; 2}/(2c_{\; 1}^{\; 
2})-1) (1-v^{\; 2}/c_{\; 1}^{\; 2})^{\; -1/2}
\times \\ \\ \ln ((x-vt)^{\; 2}+(1-v^{\; 2}/c_{\; 1}^{\; 2})y^{\; 2})
+(1-v^{\; 2}/c_{\; 2}^{\; 2})^{\; 1/2}
\ln ((x-vt)^{\; 2}+\\ \\(1- v^{\; 2}/c_{\; 2}^{\; 2}) y^{\; 2})
].
\end{array}
$$
Here $ c_{\; 1}=(\mu /\rho )^{\; 1/2}$ 
 is the speed of transversal sound waves, $c_{\; 2}=[(\lambda +2\mu )/\rho 
]^{\; 1/2}$ 
 is the speed of longitudinal sound waves, $\lambda$ and $\mu$ are Lame 
constants. 
Such relations already are not Lorentz-covariant in traditional sense.  
Formulae of the dislocation  theory can be Lorentz-covariant only 
in some special cases: for example, in case of straight dislocation
    with a Burgers vector parallel to a dislocation  line in isotropic solid. 
Therefore under certain conditions  Lorentz invariance violation in offered 
model is possible.
 May be, it will allow to explain occurrence in the last years of a number of
 field theories, not satisfying to a condition of Lorentz invariance.

Thus, in the present paper model, describing electromagnetic and gravitational
 properties of electrons and positrons, is offered.
The inclusion  of strong and weak interactions in the model is, apparently, 
future problem.
 But it already has allowed to unify advantages of soliton  models and theories
 with discrete (quantized) space-time by removing at the same time their lacks.
 Model explains quantization of electrical charge: it is proportional to the
 topological charge of dislocation, and this charge accepts quantized values 
because of discrete structure of the 4-dimensional lattice.
 At the same time existence of the lattice allows to avoid occurrence of 
divergences, in particular, results in finite mass of a particle. 
Within the framework of the given model the phenomenon of annihilation of  
particle-antiparticle pair is easily explained. The similar process is   
annihilation of dislocation pairs in solids. From topological reasons follows, 
that at a meeting of two dislocations with Burgers vectors, which are equal in 
magnitude and opposite in direction, ideal structure of a lattice restore.
 Both 
dislocations  (that is both particles) thus disappear, and their energy is 
radiated as elastic waves. As follows from above-stated, the waves of 
antisymmetric distortions are perceived by us as electromagnetic, and 
symmetric  as gravitational. It is important to note, that the relativistic 
properties of particles occur in this model as consequences of 
the classical Newton mechanics of a 4-dimensional deformable solid.

\begin {center}

ACKNOWLEDGEMENTS
\end {center}

Useful discussions with I.V. Barashenkov, M.B. Mineev-Weinstein and
O.K. Pashaev are kindly acknowledged.

\begin {center}

REFERENCES
\end {center}

\parindent=0cm

[1] D.K. Ross, ``Planck's Constant, Torsion, and Space-Time Defects", {\it Int. J. Theor. Phys.}, 
{\bf 28}, 1333-1340 (1989).

[2] A. Unzicker, ``Teleparallel space--time with torsion yields geometrization of
 electrodynamics with quantized charges." 
preprint gr-qc/9612061

[3] F.W. Hehl, P. von der Heyde, G.D. Kerlick, J.M. Nester, ``General relativity
with spin and torsion: Foundations and prospects." {\it Rev. 
Mod. Phys.}, {\bf 48}, 393-416 (1976).

[4] A.M. Kosevich, ``The dynamical theory of dislocations." {\it UFN}, {\bf 84},  579-609 
[~{\it Sov. Phys. Uspekhi}, {\bf 7}, 837-854~] (1964). 

[5] A.I. Musienko, V.A. Koptsik, ``A New Variant of the Gauge Theory
of Linear Defects in Crystals." {\it Kristallografija}, {\bf 40}, 
438-445 
 [~{\it Crystallography Reports},  {\bf 40}, 398-405~] (1995). 

[6] A.I. Moussienko, V.A. Koptsik, ``The Gauge Theory of Dislocations and
Disclinations in Crystals with Multiatomic Lattices." {\it Kristallografija}, {\bf 41}, 
586-590
 [~{\it Crystallography Reports}, {\bf 41}, 550-554~]  (1996).  

[7] F.W. Hehl, J.D. McCrea, ``Bianchi identities and the automatic conservation
of energy-momentum and angular momentum in general-relativistic field
theories." {\it Found. Phys.}, {\bf 16}, 267-293 (1986). 

[8] M.O. Katanaev, I.V. Volovich, ``Theory of defects in solids and three-dimensional gravity." {\it Ann. Phys. (N.Y.)}, {\bf 216},
 1-28 (1992). 

[9] A. Einstein, ``Einheitliche Feldtheorie von Gravitation und
Elektrizit\"at." {\it Sitzunsber. Preuss. Akad. Wiss., phys.-math.
 Kl.}, 414-419  (1925).  [ There is a Russian translation: A. Einstein, 
{\it Collection of 
Scientific Papers}, Vol. {\bf 2}, Nauka, Moscow, 171-177 (1966)]. 

[10] J.P. Hirth and J. Lothe, {\it Theory of Dislocations}, (McGraw-Hill,
 New York, 1968). 

[11] S. Coleman, ``Quantum sine-Gordon equation as the 
massive Thirring model." {\it Phys. Rev. D}, {\bf 11}, 
2088-2097 (1975).

[12] R. Rajaraman, {\it Solitons and instantons}, (North-Holland, 
Amsterdam, 1982). 

[13] J. Weertman, J.R. Weertman, ``Moving Dislocations." In: {\it Dislocations in Solids},
F.R.N.~Nabarro, ed., Vol. {\bf 3}, Moving Dislocations. 
(North-Holland, Amsterdam, 1980). p. 1-59.

\newpage
\begin{center}

FIGURE
\end{center}

FIG. 1. Dislocation in (2+1)-dimensional discrete space-time with a Burgers 
vector, parallel to an axis of time. The dislocation line on the figure is 
parallel to a Burgers vector.

\end{document}